\documentclass[pre,twocolumn,final,superscriptaddress,showpacs]{revtex4}
\usepackage{amsmath}
\usepackage{amsfonts}
\usepackage{amssymb}
\usepackage{graphicx}
\usepackage{times}
\usepackage{txfonts}
\usepackage{textcomp}
\usepackage{bm}
\usepackage{euscript}
\usepackage[colorlinks, bookmarks=false, citecolor=blue, linkcolor=red, urlcolor=blue]{hyperref}
\usepackage{bibentry}

\newcommand{\ignore}[1]{}
\newcommand{\nobibentry}[1]{{\let\nocite\ignore\bibentry{#1}}}

\begin{document}
\nobibliography*

\title{Finite Size Scaling in the Kuramoto Model}
\author{Tommaso Coletta}
\affiliation{School of Engineering, University of Applied Sciences of Western Switzerland, CH-1951 Sion, Switzerland}
\author{Robin Delabays}
\affiliation{School of Engineering, University of Applied Sciences of Western Switzerland, CH-1951 Sion, Switzerland}
\affiliation{Section de Math\'ematiques, Universit\'e de Gen\`eve, CH-1211 Gen\`eve, Switzerland}
\author{Philippe Jacquod}
\affiliation{School of Engineering, University of Applied Sciences of Western Switzerland, CH-1951 Sion, Switzerland}
\date{\today}

\begin{abstract}
We investigate the scaling properties of the order parameter and the largest nonvanishing Lyapunov exponent for the fully locked 
state in the Kuramoto model with a finite number $N$ of oscillators. 
We show that, for any finite value of $N$, both quantities scale as $(K-K_L)^{1/2}$ with the coupling strength $K$ 
sufficiently close to the locking threshold $K_L$.
We confirm numerically these predictions for oscillator frequencies evenly spaced in the interval $[-1, 1]$ and additionally find 
that the coupling range $\delta K$ over which this scaling is valid shrinks like 
$\delta K \sim N^{-\alpha}$ with $\alpha\approx1.5$ as $N \rightarrow \infty$. 
Away from this interval, the order parameter exhibits the infinite-$N$ behavior $r-r_L \sim (K-K_L)^{2/3}$ proposed by 
Paz\'o [\nobibentry{Pazo2005}].
We argue that the crossover between the two behaviors occurs
because at the locking threshold, the upper bound of the continuous part of the 
spectrum of the fully locked state approaches zero as $N$ increases. Our results clarify the convergence to the $N \rightarrow \infty$ limit in the 
Kuramoto model.
\end{abstract}


\maketitle
\section{Introduction}
The coupled oscillator model introduced by Kuramoto in the late 70's has
established itself as a paradigmatic model for the study of synchronization phenomena,
where it opened a vast area of research.
The Kuramoto model allows to investigate the interplay between the tendency that individual oscillators have to
run at their natural frequency and a sinusoidal all-to-all coupling which attempts to synchronize the oscillators \cite{Kuramoto1975,Kuramoto1984}.
Kuramoto elegantly solved the
model in the limit of infinitely many oscillators with natural frequencies drawn from a Lorentzian distribution, for which
he showed that upon increasing the coupling between oscillators, the system undergoes a transition from
an incoherent disordered phase to a partially synchronized state with a finite fraction of oscillators rotating in unison \cite{Kuramoto1975,Kuramoto1984,Strogatz01}.

Several evolutions of the original Kuramoto model have been investigated, including
models with different natural frequency distributions, with couplings defined on a complex network topology, oscillators 
with inertia, couplings with frustration, with time delays and even negative couplings to 
name but a few \cite{Acebron2005,Arenas2008}. 
Such extensions are motivated by the connection that the Kuramoto model has with several physical systems, ranging from synchronization phenomena in biological systems
\cite{Winfree1967,Ermentrout1991} to Josephson junction arrays \cite{Wiesenfeld96}, via synchronous AC electric power systems
\cite{Dorfler2013,Motter2013,Coletta2016}.

Recently, there has been a renewed interest in the finite size behavior of 
the Kuramoto model \cite{Mirollo2005,Aeyels2004,Jadbabaie2004,Pazo2005,Ottino2016}. 
The problem is of interest, because all physically relevant systems and numerical simulations
deal with a finite number $N$ of oscillators, which makes finding solutions to the Kuramoto model mathematically more involved. 
In particular, at finite $N$, the continuum limit breaks down so that 
self consistent equations for physically relevant quantities such as the order parameter can no longer be written in a mathematically convenient integral form.
Important steps forward in the understanding of the finite size Kuramoto model include the description of the Lyapunov spectrum 
for the fully locked state as well as estimates of the critical coupling necessary for synchronization to occur \cite{Mirollo2005,Aeyels2004,Jadbabaie2004}. A complete understanding of the transition to the infinite-$N$ behavior is however still lacking.

In this work we consider the finite size Kuramoto model [\onlinecite{Kuramoto1975,Kuramoto1984}] on 
the complete graph $\mathcal G$ with $N$ nodes and $|\mathcal E|=N(N-1)/2$ edges
\begin{equation}\label{eq:Kuramoto model}
 \dot{\theta_i}=\omega_i+\frac{K}{N}\sum_{j=1}^N\sin\left(\theta_j-\theta_i\right)\,,  \quad i=1,\dots,N \, , 
\end{equation}
where $\theta_i$ and $\omega_i$  are the phases and the natural frequencies of the oscillators, respectively,
and $K/N>0$ is the coupling strength. 
For natural frequencies defined on a bounded interval, there exists a critical value of the coupling $K_L^N$ for which the  
system is in a fully locked state where all oscillators synchronize, 
with $K_L^N\rightarrow K_L^\infty$ as $N\rightarrow\infty$ \cite{Strogatz01,vanHemmen1993,Dorfler11,Verwoerd11}.
For the particular case of uniformly distributed frequencies, the main focus of this work, it has been found that 
the transition from the incoherent state to full synchrony is of first order \cite{Pazo2005}. 
We investigate the scaling properties of the Lyapunov spectrum characterizing the linear stability of the fully locked state, 
and of the order parameter introduced by Kuramoto \cite{Kuramoto1975,Kuramoto1984}.
We show that above the locking threshold the largest non vanishing Lyapunov exponent $\lambda_2$ scales like 
$\lambda_2\sim(K-K_L)^{1/2}$. Relating the expression for the order parameter $r$ [Eq.~(\ref{eq:order_parameter}) below]
to the Lyapunov exponents, we show
that the order parameter also scales as $r-r_L\sim(K-K_L)^{1/2}$, $r_L \equiv r(K_L)$ being the 
order parameter at the locking threshold.
We confirm numerically these results for uniformly distributed 
oscillator frequencies. At first glance, our results disagree with 
Paz\'o who obtained $r-r_L\sim(K-K_L)^{2/3}$ \cite{Pazo2005}. 
The two results can be reconciled once one realizes that Paz\'o's calculation is strictly valid 
for an infinite number of oscillators only, while our results are derived for finite $N$. 
We find numerically that our finite $N$ result is always valid close enough to $K_L$. 
However, its range of validity $\delta K$ becomes narrower and narrower as $N$ increases, 
with numerical data consistent with $\delta K \sim N^{-\alpha}\,,\alpha\approx1.5$.
We further argue that the crossover is triggered by 
the dependence on $N$ of the next largest non vanishing Lyapunov exponent $\lambda_3$ at $K_L$,
$\lambda_3(K_L) \sim N^{-1/2}$. Corrections to our results being of order $\lambda_3^{-1}$,
they can no longer be neglected as $N\rightarrow\infty$.
A side result of our approach is that all Lyapunov exponents of the fully locked state of the Kuramoto model are monotonically decreasing functions 
of the coupling strength. This directly implies that the linear stability of the fully locked state improves as the oscillator coupling
is increased and that if the locked state exists at $K_0$, it exists at all coupling strengths $K \ge K_0$. We note that 
this result could have been anticipated starting from the properties of the Lyapunov spectrum discussed 
in Ref.~\cite{Mirollo2005}.

This paper is organized as follows. Section \ref{sec:Definitions} recalls the definition of fully locked states
in the Kuramoto model. Sections \ref{sec:Monotonicity r} and \ref{sec:Monotonicity lambda} present the 
calculation of the monotonicity of the Lyapunov exponents as a function of the coupling constant.
Section \ref{sec:Finite size corrections} discusses the behavior of the largest nonvanishing Lyapunov exponent and of the order parameter in the immediate
vicinity of the phase-locking threshold for a large but finite number of oscillators.

\section{The Kuramoto model}\label{sec:Definitions}
We consider the Kuramoto model defined by Eq.~(\ref{eq:Kuramoto model}) 
and $\omega_i\in[-1,1]$ though our results remain valid
for distributions defined on bounded intervals.
Introducing the order parameter \cite{Kuramoto1975,Kuramoto1984}
\begin{equation}\label{eq:order_parameter}
 re^{i\psi}=\frac{1}{N}\sum_{i=1}^{N}e^{i\theta_i}\,,
\end{equation}
Eq.~(\ref{eq:Kuramoto model}) can be rewritten as
\begin{equation}
\dot{\theta_i}=\omega_i+Kr\sin\left(\psi-\theta_i\right)\,, \quad i=1,\dots,N.
\end{equation}
Given the invariance of the Kuramoto model under a global shift of all phases, we can set $\psi=0$.
Without loss of generality we consider natural frequencies such that $\sum_i\omega_i=0$, which is tantamount 
to considering the system in a rotating frame.
For $K>K_L$, Eq.~(\ref{eq:Kuramoto model}) admits
stationary solutions $\{\theta_i^{(0)}\}$ given by
\begin{equation}\label{eq:locked states}
 \sin\left(\theta_i^{(0)}\right)=\frac{\omega_i}{Kr}\,, \quad i=1,\dots,N\,.
\end{equation}
They are referred to as {\it fully locked states}.
The linear stability of fully locked states is governed by the spectrum of the 
stability matrix $\bm{M}$, obtained by linearizing Eq.~(\ref{eq:Kuramoto model}) 
close to $\{\theta_{i}^{(0)}\}$, and defined as
\begin{equation}\label{eq:stability matrix}
 M_{ij}=
\left\{
 \begin{array}{cl}
 \displaystyle \frac{K}{N}\cos\left(\theta_j^{(0)}-\theta_i^{(0)}\right)\,, & i\neq j\,,\\[3mm]
 \displaystyle -\frac{K}{N}\sum_{l\neq i}\cos\left(\theta_l^{(0)}-\theta_i^{(0)}\right)\,, & i=j\,.
 \end{array}\right.
 \end{equation}

Since the stationary solutions of Eq.~(\ref{eq:Kuramoto model}) are invariant under a global rotation of all angles, 
one of the eigenvalues of $\bm M$ is identical to zero.
A stationary solution $\{\theta^{(0)}_i\}$ is linearly stable as long as $\bm M$ is negative semidefinite.
This condition ensures that for any small perturbation around $\{\theta^{(0)}_i\}$, the system's state, subject to
the dynamics of Eq.~(\ref{eq:Kuramoto model}), returns to $\{\theta^{(0)}_i\}$ exponentially fast.
The eigenvalues $\lambda_i$ of $\bm M$ are referred to as
the Lyapunov exponents and the linear stability condition is expressed as
\begin{equation}\label{eq:eigenvalues stability matrix}
 \lambda_1=0>\lambda_2\geq\lambda_3\dots\geq\lambda_N\,.
\end{equation}
In what follows $\{{\bm u}^{(q)}\}$, $q=1,\ldots,N$ is the othonormal basis of eigenvectors of $\bm M$ defined by 
$\bm M{\bm u}^{(q)}=\lambda_q{\bm u}^{(q)}$.
In particular $\bm{u}^{(1)}=(1,\dots,1)/\sqrt{N}$ is the eigenvector associated with $\lambda_1=0$.

According to Sylvester's criterion, a necessary condition for $\bm{M}$
to be negative semidefinite is that all its diagonal elements are negative (i.e. $M_{ii}\leq0$ 
for all $i$). This implies [\onlinecite{Mirollo2005}]
\begin{equation}
\begin{array}{c}
\displaystyle -\frac{K}{N}\sum_{l\neq i}\cos\left(\theta_l^{(0)}-\theta_i^{(0)}\right)=
\displaystyle -\frac{K}{N}\left[\sum_{l=1}^N\cos\left(\theta_l^{(0)}-\theta_i^{(0)}\right)-1\right]\leq0\\[3mm]
\displaystyle \Rightarrow -Kr\cos\left(\theta_i^{(0)}\right)+\frac{K}{N}\leq0  \\[3mm]
\displaystyle \Rightarrow 0\leq\frac{1}{rN}\leq \cos\left(\theta_i^{(0)}\right) \, ,  \quad \forall i=1, \ldots N \,.
\end{array}
\end{equation}
The positivity of the cosine, together with Eq.~(\ref{eq:locked states}), allows to rewrite 
\begin{equation}\label{eq:cosine fully locked}
 \cos\left(\theta_i^{(0)}\right)=\sqrt{1-(\omega_i/Kr)^2} \, , \quad \forall i=1, \ldots N\,.
\end{equation}
This choice actually corresponds to the unique stable locked state solution of the all-to-all Kuramoto model \cite{Mirollo2005,Aeyels2004}.

\section{Monotonicity of the order parameter}\label{sec:Monotonicity r}
In this section we show that for the stable fully locked state the magnitude of the order 
parameter $r$ grows monotonically as the coupling constant $K$ increases.
This result has already been reported  in the literature [\onlinecite{Jadbabaie2004, Mirollo2005}], however
our calculation below is based on a novel formalism which we will use later on. We therefore 
present it.

We start by expressing the square of the modulus of the order parameter as 
\begin{equation}\label{eq:square of order param}
 r^2=\frac{1}{N^2}\left[N+2\sum_{j>i}\cos\left(\theta_j^{(0)}-\theta_i^{(0)}\right)\right]\,,
\end{equation}
where the sum runs over all pairs of oscillators.
Taking the derivative of Eq.~(\ref{eq:square of order param}) with respect to $K$
gives
\begin{equation}\label{eq:derivative order parameter}
 \frac{dr}{dK}=-\frac{1}{rN^2}\sum_{j>i}
 \sin\left(\theta_j^{(0)}-\theta_i^{(0)}\right)\frac{d}{dK}\left(\theta_j^{(0)}-\theta_i^{(0)}\right)\,.
\end{equation}
To obtain an expression for $d\left.\left(\theta_j^{(0)}-\theta_i^{(0)}\right)\right/dK$, we take the derivative of
the stationary condition
\begin{equation}
 0=\omega_i+\frac{K}{N}\sum_{j=1}^N\sin\left(\theta_j^{(0)}-\theta_i^{(0)}\right)
\end{equation}
with respect to $K$.
This gives
\begin{equation}\label{eq:Angle derivative}
\begin{array}{c}
\displaystyle -\sum_{j=1}^N\sin\left(\theta^{(0)}_j-\theta^{(0)}_i\right)
=K\sum_{j=1}^N\cos\left(\theta^{(0)}_j-\theta^{(0)}_i\right)\frac{d}{dK}\left(\theta_j^{(0)}-\theta_i^{(0)}\right)\\[3mm]
\Rightarrow\displaystyle \bm{\omega}/K={\bm M}\frac{d}{dK}{\bm{\theta}}^{(0)}\,,
\end{array}
\end{equation}
where ${\bm \theta}^{(0)}=(\theta^{(0)}_1,\dots,\theta^{(0)}_N)$ and 
${\bm \omega}=(\omega_1,\dots,\omega_N)$. Since $\bm M$ is singular, we invert Eq.~(\ref{eq:Angle derivative}) using the 
Moore-Penrose pseudoinverse of $\bm M$ defined as
\begin{equation}\label{eq:Definition pseudo inverse}
  {\bm M}^{-1} ={\bm T}\left(
 \begin{array}{cccc}
 0& & & \\
  &\lambda_2^{-1}&& \\
  & &\ddots& \\
  & & &\lambda_N^{-1}
 \end{array}\right) {\bm T}^\top \,,
\end{equation}
where ${\bm T}=(\bm{u}^{(1)},\ldots,\bm{u}^{(N)})$ and ${\bm M}^{-1}{\bm M}={\bm M}{\bm M}^{-1}=\mathbb{I}-\bm{u}^{(1)}{\bm{u}^{(1)}}^\top $.
Multiplying Eq.~(\ref{eq:Angle derivative}) by ${\bm M}^{-1}$ yields
\begin{equation}\label{eq:dthetadk}
\frac{d}{dK}{\bm\theta}^{(0)}=
 {\bm M}^{-1}\frac{\bm \omega}{K}+
  \frac{1}{N}\frac{d}{dK}
  \left(\begin{array}{c}
  \sum_l\theta_l^{(0)}\\
  \vdots \\
  \sum_l\theta_l^{(0)}\\
 \end{array}\right)\,.
\end{equation}
Finally, the difference between any two components of the expression above is given by
\begin{equation}\label{eq:delta theta 2}
\begin{array}{lll}
\displaystyle \frac{d}{dK}\left(\theta_j^{(0)}-\theta_i^{(0)}\right)&=&\displaystyle\frac{1}{K}\sum_k\left(M^{-1}_{jk}-M^{-1}_{ik}\right)\omega_k\\
  &=&\displaystyle \frac{1}{K}\sum_{k,\,l\geq2}\left(u_j^{(l)}-u_i^{(l)}\right)\frac{1}{\lambda_l}u_k^{(l)}\omega_k \,,
\end{array}
\end{equation}
where the terms with $\sum_l\theta_l^{(0)}$ in Eq.~(\ref{eq:dthetadk}) drop due to the global rotational invariance of the Kuramoto model.
Injecting this result into Eq.~(\ref{eq:derivative order parameter}) gives

\begin{equation}\label{eq:derivative order parameter2}
 \frac{dr}{dK}=-\frac{1}{rKN^2}
  \sum_{\substack{j>i \\ k,\,l\geq2}}
 \frac{1}{\lambda_l}\sin\left(\theta_j^{(0)}-\theta_i^{(0)}\right)\left(u_j^{(l)}-u_i^{(l)}\right)u_k^{(l)}\omega_k\,.
\end{equation}

In order to determine the sign of the right-hand side of Eq.~(\ref{eq:derivative order parameter2}) 
it is useful to introduce the incidence matrix $\bm B$ of the network. 
Given a graph $\mathcal{G}$ of $N$ nodes and $|\mathcal{E}|$ edges and given an 
arbitrary orientation of each edge, the incidence matrix $\bm B\in\rm I\!R^{N\times|\mathcal{E}|}$ is defined as follows
\begin{equation}
\begin{array}{ccc}
 B_{il}=\left\{
 \begin{array}{cl}
  1\,, \quad &\textrm{if }i\textrm{ is the source of edge }l \,,\\
 -1\,, \quad &\textrm{if }i\textrm{ is the sink of edge }l \,,\\
  0\,, \quad &\textrm{otherwise} \,.\\
  \end{array}
 \right.
\end{array}
\end{equation}
The product ${\bm B}^\top{\bm \theta}^{(0)}$ is a vector in $\rm I\!R^{|\mathcal{E}|}$ whose $l^\textrm{th}$ entry is equal to
$\theta_i^{(0)}-\theta_j^{(0)}$ where $i$ and $j$ are the nodes connected by edge $l$, 
and where the sign of this difference depends on the arbitrary choice of orientation of the edge
($i$ is the source and $j$ is the sink in this case).
Similarly, given a vector $\bm v\in {\rm I\!R} ^{|\mathcal{E}|}$, the product $\bm B{\bm v}$ is a vector in $\rm I\!R^N$ 
whose $i^\textrm{th}$ entry is equal to the sum $\sum_{l}\pm v_{l}$ over all edges $l$ connected to node $i$, 
and with the sign $\pm$ fixed by the nature (sink or source) of site $i$. 

We then rewrite the Kuramoto model, Eq.~(\ref{eq:Kuramoto model}), in vector form using the incidence 
matrix we just introduced 
\begin{equation}
 \dot{\bm\theta}={\bm \omega}-\frac{K}{N}{\bm B}\cdot\bm{\mathrm{sin}}\left({\bm B}^T\bm\theta\right)\,,  
\end{equation}
where we defined $\bm{\mathrm{sin}}({\bm x})\equiv(\sin(x_1),\dots,\sin(x_{|\mathcal{E}|}))$ 
for $\bm x \in \rm I\!R^{|\mathcal{E}|}$.
Thus, for a stationary solution we have
\begin{equation}\label{eq:Stationary sol 2}
 {\bm \omega}=\frac{K}{N}{\bm B}\cdot\bm{\mathrm{sin}}\left({\bm B}^T\bm\theta^{(0)}\right)\,.
\end{equation}
This compact formulation allows to write
\begin{equation}\label{eq:Identity}
\begin{array}{lll}
\displaystyle\sum_k u_k^{(l)}\omega_k
                   &=&\displaystyle \frac{K}{N}\left({\bm B}^\top{\bm u}^{(l)}\right)^T \cdot\bm{\mathrm{sin}}\left({\bm B}^T\bm\theta^{(0)}\right) \\[3mm]
                   &=&\displaystyle \frac{K}{N}\sum_{j>i}\left(u_i^{(l)}-u_j^{(l)}\right)\sin\left(\theta^{(0)}_i-\theta^{(0)}_j\right)\,. 
\end{array}
\end{equation}
Injecting this last identity into Eq.~(\ref{eq:derivative order parameter2})
gives for the fully locked state
\begin{equation}\label{eq:derivative order parameter3}
 \frac{dr}{dK}=-\frac{1}{rK^2N}
  \sum_{l\geq2}
 \frac{1}{\lambda_l}\left(\sum_k{u_k^{(l)}\omega_k}\right)^2\geq0\,,
\end{equation}
since $r\geq0$ and $\lambda_l<0$ for $l\geq2$ in the stable fully locked state.
The order parameter is therefore a monotonously increasing function of $K$.

\section{Monotonicity of the Lyapunov exponents}\label{sec:Monotonicity lambda}
Next, given the stable fully locked state $\{\theta^{(0)}_i\}$, we compute the variation of its Lyapunov exponents as a function of the coupling strength $K$.
Because ${\bm M}$ is real and symmetric, we can apply the Hellmann-Feynmann theorem \cite{Cohen1977} to calculate $d\lambda_q/dK$.
We obtain
\begin{equation}\label{eq:HF theorem}
 \frac{d\lambda_q}{dK}={\bm{u}^{(q)}}^\top \frac{d {\bm M}}{dK} \bm{u}^{(q)}\,.
\end{equation}
We express the derivative of the stability matrix with respect to $K$ as $d{\bm M}/dK={\bm M}/K+\bar{{\bm M}}$, which injected back into
Eq.~(\ref{eq:HF theorem}) yields
\begin{equation}\label{eq:dLambda/dk}
 \frac{d\lambda_q}{dK}=\frac{\lambda_q}{K}+ {\bm{u}^{(q)}}^\top \bar{{\bm M}} \bm{u}^{(q)}\, .
\end{equation}
The matrix $\bar{{\bm M}}$ is defined by
\begin{equation}\label{eq:Bar M}
 \bar{M}_{ij}=\left\{
 \begin{array}{cl}
 \displaystyle -\frac{K}{N}\sin\left(\theta_j^{(0)}-\theta_i^{(0)}\right)\frac{d}{dK}\left(\theta_j^{(0)}-\theta_i^{(0)}\right)\,, & i\neq j\,,\\[3mm]
 \displaystyle \frac{K}{N}\sum_{l\neq i}\sin\left(\theta_l^{(0)}-\theta_i^{(0)}\right)\frac{d}{dK}\left(\theta_j^{(0)}-\theta_i^{(0)}\right)\,, & i=j\,.
 \end{array}\right.
\end{equation}

Next we show that for the linearly stable fully locked state, $d\lambda_q/dK\leq0$ for all values of $q$, i.e.
the Lyapunov exponents are monotonically decreasing functions of the coupling strength.
If the stationary solution considered is linearly stable 
(i.e. $\lambda_i\leq0\,,\forall i$), the first term in the right-hand side of 
Eq.~(\ref{eq:dLambda/dk}) is negative and only the sign of the second term needs to be determined.
We note that ${\bar {\bm M}}$ shares the same zero row sum property as ${\bm M}$, thus ${\bm{u}^{(1)}}^\top \bar{{\bm M}} \bm{u}^{(1)}=0$ and one
readily obtains that $d\lambda_1/dK=0$ as should be.

Using Eqs.~(\ref{eq:locked states}) and (\ref{eq:cosine fully locked}), 
and expanding $\sin\left(\theta_j^{(0)}-\theta_i^{(0)}\right) = \sin \theta_j^{(0)} \, \cos \theta_i^{(0)} -
\cos \theta_j^{(0)}\, \sin \theta_i^{(0)}$ we obtain
\begin{equation}\label{eq:Sine of difference}
\begin{array}{l}
 \sin\left(\theta_j^{(0)}-\theta_i^{(0)}\right)=\frac{\sqrt{(Kr)^2-\omega_i^2}\sqrt{(Kr)^2-\omega_j^2}}{(Kr)^2}
 \left(\frac{\omega_j}{\sqrt{(Kr)^2-\omega_j^2}}-\frac{\omega_i}{\sqrt{(Kr)^2-\omega_i^2}}\right)\,,
\end{array}
\end{equation}
as well as 
\begin{equation}
\begin{array}{c}
 \displaystyle\frac{d}{dK}\left(\theta_j^{(0)}-\theta_i^{(0)}\right)=\displaystyle\frac{d}{dK}\left[\arcsin\left(\frac{\omega_j}{Kr}\right)
 -\arcsin\left(\frac{\omega_i}{Kr}\right)\right] \\[3mm]
 =\displaystyle-\frac{1}{Kr}\left(r+K\frac{dr}{dK}\right)
 \left(\frac{\omega_j}{\sqrt{(Kr)^2-\omega_j^2}}-\frac{\omega_i}{\sqrt{(Kr)^2-\omega_i^2}}\right)\,.
\end{array} 
\end{equation}
Hence, the product
\begin{equation}\label{eq:sign off diag elements}
\begin{array}{c}
\displaystyle -\frac{K}{N}\sin\left(\theta_j^{(0)}-\theta_i^{(0)}\right)\frac{d}{dK}\left(\theta_j^{(0)}-\theta_i^{(0)}\right)\\[3mm]
 =\frac{K}{N}\frac{\sqrt{(Kr)^2-\omega_i^2}\sqrt{(Kr)^2-\omega_j^2}}{(Kr)^3}\left(r+K\frac{dr}{dK}\right)\left(\frac{\omega_j}{\sqrt{(Kr)^2-\omega_j^2}}-\frac{\omega_i}{\sqrt{(Kr)^2-\omega_i^2}}\right)^2\,,
\end{array}
\end{equation}
is positive, since $dr/dK\geq0$ as shown in Section \ref{sec:Monotonicity r}. This result implies that for the Kuramoto model on the complete
graph, increasing the coupling strength systematically reduces the difference $|\theta_{i}^{(0)}-\theta_{j}^{(0)}|$ for all pairs of 
oscillators $i$ and $j$.

Putting all this together, $\bar{{\bm M}}$ is a zero row sum matrix and Eq.~(\ref{eq:sign off diag elements}) proves that all its 
off diagonal entries are positive. Thus, invoking Gershgorin's circle theorem \cite{Horn_MatrixAnalysis}, we conclude that $\bar{{\bm M}}$ is negative semidefinite
and thus ${\bm{u}^{(q)}}^\top \bar{{\bm M}} \bm{u}^{(q)}\leq0$.
This concludes the proof that the Lyapunov exponents of the fully locked solution of the
Kuramoto model are decreasing functions of the coupling, i.e.
\begin{equation}\label{eq:dlambdadK}
 \begin{array}{l}
 \displaystyle \frac{d\lambda_1}{dK}=0\,, \\[3mm]
 \displaystyle \frac{d\lambda_q}{dK}<0\,,\quad 2\leq q\leq N\,.
 \end{array}
\end{equation}
This result implies that if the fully locked state is stable at $K_0\in[0,+\infty)$, 
then this solution remains linearly stable and thus can be continuously 
followed in the interval $K_0\leq K\leq+\infty$. 
In other words, starting from a stable configuration, $d\lambda_q/dK\leq0$
for all $q$ ensures that no instability occurs as the coupling increases 
(i.e. that none of the Lyapunov exponents, except $\lambda_1$, vanishes).
Equation (\ref{eq:dlambdadK}) not only implies that the stable fully locked state remains stable as the coupling strength is increased, 
but also that it becomes ``more`` stable, in the sense that more negative Lyapunov exponents correspond to
shorter timescales to return to equilibrium.
We note that the monotonicity of the Lyapunov exponents with the coupling can also be derived starting from the properties of the 
spectrum of the fully locked state presented in Ref. [\onlinecite{Mirollo2005}].

\section{Scaling behavior of the order parameter}\label{sec:Finite size corrections}
It is known that for uniformly distributed oscillator frequencies,
the transition between the incoherent and the fully synchronized state is of first order \cite{Pazo2005}.
For finite $N$, this transition occurs as the coupling is increased above $K_L^N$
and is characterized by a discontinuous jump in the order parameter from $0$ to $r_L^N$.
The values of 
$r_L^N$ and $K_L^N$ depend explicitly on the number of oscillators and can be calculated for specific distributions of natural
frequencies \cite{Ottino2016}.

The investigation of fully locked states in the infinite-$N$ version of the Kuramoto model with frequency distributions 
supported on a bounded interval dates back to Ermentrout [\onlinecite{Ermentrout1985}] who
showed that for uniform frequency distributions the locking threshold and the order parameter
at the locking transition are given by $K_{L}^\infty=4/\pi$ and $r_L^\infty=\pi/4$. 
More recently, Paz\'o [\onlinecite{Pazo2005}] showed that for the infinite-$N$ Kuramoto model and a 
uniform box distribution of natural frequencies, $[-1,1]$ 
the order parameter above the locking threshold scales like 
\begin{equation}
 r-r_L^\infty=\left(\frac{9\pi^7}{2^{17}}\right)^{1/3}(K-K_L)^{2/3}+\mathcal{O}(K-K_L)\,.
\end{equation}

We next show analytically that for the finite size Kuramoto model and uniform frequency distribution, 
the scaling of the order parameter instead goes like $(r-r_L^N) \sim (K-K_L^N)^{1/2}$.
Furthermore, we find numerically that the range of validity of this behavior decreases with $N$. 

We start off from Eq.~(\ref{eq:dLambda/dk}) for $d\lambda_q/dK$, and express the average 
${\bm{u}^{(q)}}^\top \bar{M} \bm{u}^{(q)}$
using Eq.~(\ref{eq:delta theta 2}).
We obtain
\begin{equation}\label{eq:dLambda/dk 2}
 \frac{d\lambda_q}{dK}=\frac{\lambda_q}{K}+\sum_{l\geq2}\frac{C_l^{(q)}(K)}{\lambda_l}\,,
\end{equation}
with 
\begin{equation}\label{eq:coefficient dLambda/dk}
C_l^{(q)}=\frac{1}{N}\sum_{j>i}{\sin\left(\theta_j^{(0)}-\theta_i^{(0)}\right)\left(u_j^{(l)}-u_i^{(l)}\right)\left(u_j^{(q)}-u_i^{(q)}\right)^2\sum_k u_k^{(l)}\omega_k}\,.
\end{equation}

In Section \ref{sec:Monotonicity lambda}, we showed that the Lyapunov exponents of the stable fully locked state decrease as the coupling is reduced.
Upon reducing $K$, the locking threshold $K_L^N$ is eventually reached at which point the locked state becomes unstable and 
ceases to exist. This bifurcation is accompanied with the vanishing of the Lyapunov exponent $\lambda_2$.
Assuming that ${C_2^{(2)}(K)}$ does not vanish at $K=K_L^N$, sufficiently close to $K_L^N$, we can approximate Eq.~(\ref{eq:dLambda/dk 2}) for $q=2$  by
\begin{equation}\label{eq:dLambda/dk 3}
 \frac{d\lambda_2}{dK}\approx\frac{C_2^{(2)}(K_L^N)}{\lambda_2}\,.
\end{equation}
Solving Eq.~(\ref{eq:dLambda/dk 3}) yields
\begin{equation}\label{eq:critical scaling lambda2}
\lambda_2\approx -\sqrt{2C_2^{(2)}(K_L^N)}\sqrt{K-K_L^N}\,.
\end{equation}
This result indicates that the largest non zero Lyapunov exponent approaches zero with a square root behavior in the vicinity of 
the bifurcation. 
For symmetrically distributed natural frequencies $\omega_i$, it follows from Eq.~(8) in Ref.~\cite{Mirollo2005}, together with
Eq.~(\ref{eq:Sine of difference}) that $C_2^{(2)}(K_L^N)$ is finite.
Numerical results to be presented below for that case corroborate Eq.~(\ref{eq:critical scaling lambda2}).

Eq.~(\ref{eq:dLambda/dk 2}) also captures the asymptotic behavior of the Lyapunov exponents in the limit 
$K\rightarrow+\infty$. Since the Lyapunov exponents are decreasing functions of the coupling, 
at large values of $K$ we have $1/\lambda_l\ll1$ for all $l\geq2$.
Neglecting the second term in the right-hand side of Eq.~(\ref{eq:dLambda/dk 2}) yields 
\begin{equation}
 \lambda_q\approx -K \quad q=2,\ldots,N\,,
\end{equation}
as expected.
Since $|\theta_i^{(0)}-\theta_j^{(0)}|$ decreases with $K$
for all $i,j$, when $K\rightarrow+\infty$ the value of all cosines entering the definition of the stability matrix
Eq.~(\ref{eq:stability matrix}) approaches $1$ in which case its eigenvalues are $-K$ with multiplicity $N-1$,
and $0$ with multiplicity $1$.

We next turn our attention to the order parameter close but above locking. 
When the coupling approaches the locking threshold, $\lambda_2\rightarrow0$. This justifies to truncate the sum in
Eq.~(\ref{eq:derivative order parameter3}), keeping only the dominant term $l=2$
\begin{equation}\label{eq:dr over dK}
  \frac{dr}{dK}\approx-\frac{1}{rK^2N} \frac{1}{\lambda_2}\left(\sum_k{u_k^{(2)}\omega_k}\right)^2\,.
\end{equation}
Using the scaling behavior of $\lambda_2$ derived above, Eq.~(\ref{eq:critical scaling lambda2}), 
we obtain the leading expression for $dr/dK$ by	replacing $K$ and $r$ respectively by $K_L$ and $r_L^N$ in the right-hand side 
of Eq.~(\ref{eq:dr over dK}). Solving the resulting ordinary differential equation we obtain
\begin{equation}\label{eq:scaling rc}
 r-r_L^N\approx\frac{1}{r_L^N(K_L^N)^2N}
 \frac{2\left(\sum_k{u_k^{(2)}\omega_k}\right)^2}{\sqrt{2C_2^{(2)}(K_L^N)}}\sqrt{K-K_L^N}\,.
\end{equation}

\begin{figure}[htbp]
\centering
 \includegraphics[width=\columnwidth]{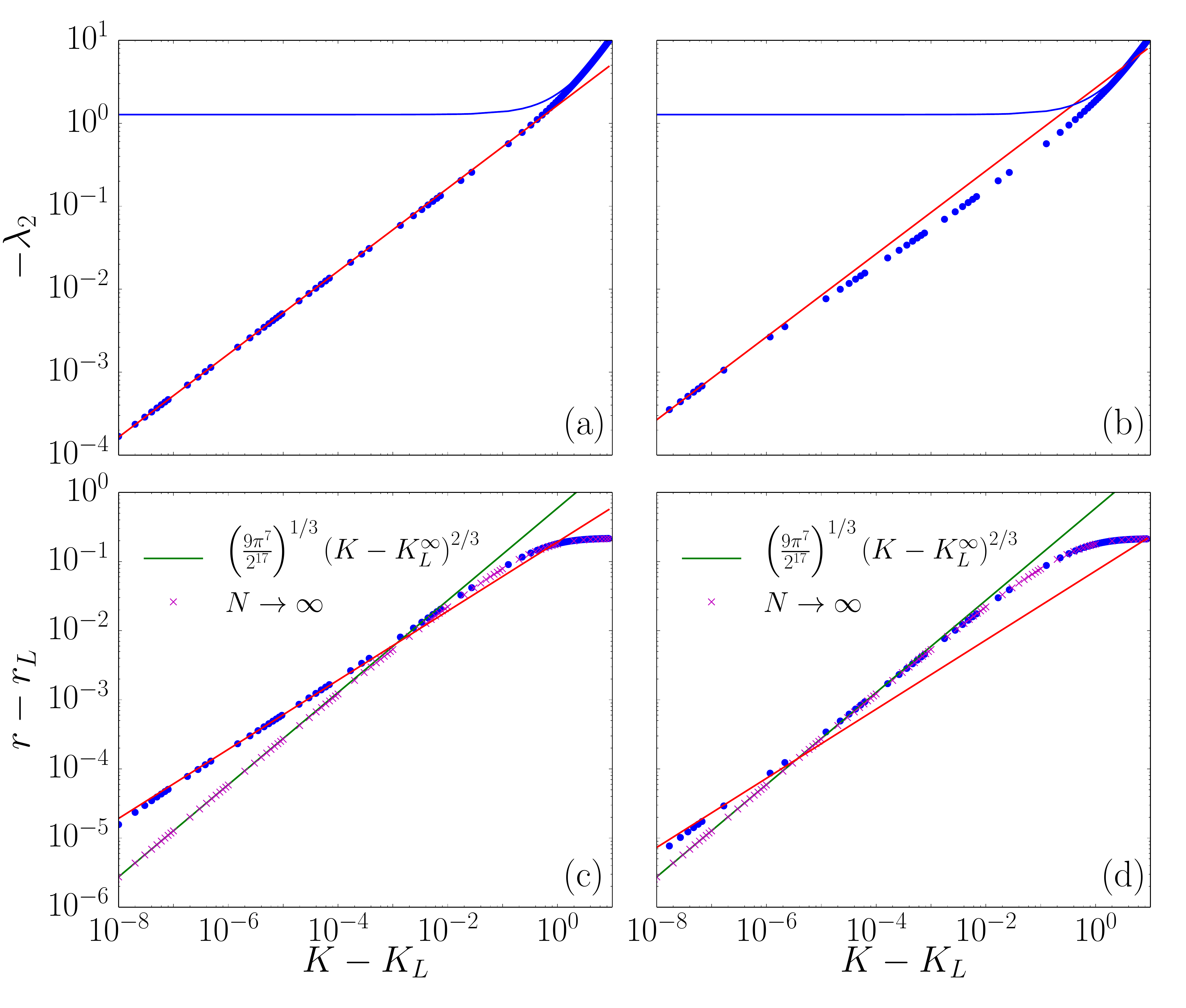}
 \caption{(Color online) $4^{th}$ order Runge-Kutta simulation results of the finite size Kuramoto model.
 Oscillator frequencies are distributed over the interval $[-1,1]$ according to the mid-point rule, Eq.~(\ref{eq:mid point rule}),
 and $N=100$ [panels a) and c)] and $5000$ [panels b) and d)]. The locking threshold $K_L^N$ is determined numerically
 with an accuracy of $10^{-9}$. The Runge-Kutta integration step is 0.005 and the maximal number of iterations is 
 $5\cdot10^7$. 
 Panels a) and b) show the square root behavior of $\lambda_2$ as a function of $K-K_L^N$. 
 The red lines give our theoretical prediction, Eq.~(\ref{eq:critical scaling lambda2}), 
 with no fitting parameter, the prefactor of Eq.~(\ref{eq:critical scaling lambda2})
 being computed numerically for the best estimate of the locking threshold obtained.
 The blue lines give the large $K$ asymptotics $\lambda_2=-K$.
 Panels c) and d) show the square root behavior of $r-r_L$ as a function of $K-K_L^N$.
 The red lines give our theoretical prediction, Eq.~(\ref{eq:scaling rc}), 
 with no fitting parameter, the prefactor of Eq.~(\ref{eq:scaling rc})
 being computed numerically for the best estimate of the locking threshold obtained.
 The plots also present the infinite-$N$ limit results for $r-r_L$ as a function of the distance $K-K_L^\infty$
 obtained by solving numerically the self consistent equation for the order parameter (crosses), 
 as well as the $2/3$ scaling exponent prediction of Ref.~[\onlinecite{Pazo2005}] (green line).}
 \label{fig:K-KL scaling mid}
\end{figure}

\begin{figure}[htbp]
\centering
  \includegraphics[width=\columnwidth]{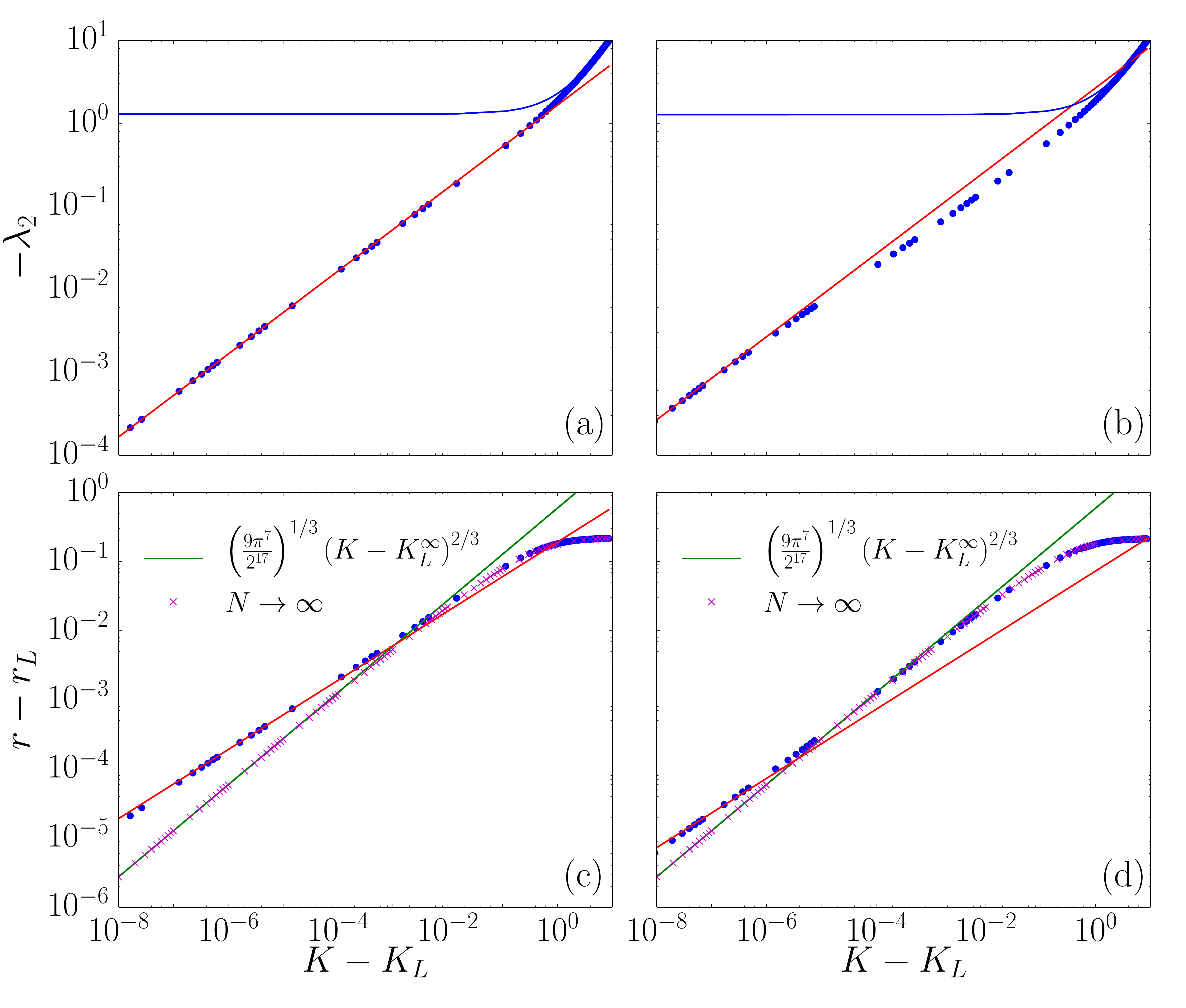}
 \caption{(Color online) Same as in Fig.~\ref{fig:K-KL scaling mid} but for oscillator frequencies 
 distributed according to the end-point rule, Eq.~(\ref{eq:end point rule}).}
 \label{fig:K-KL scaling end}
\end{figure}

To check our main results, Eqs.~(\ref{eq:critical scaling lambda2}) and (\ref{eq:scaling rc}), we numerically
simulate Kuramoto models with box distributed natural frequencies and various $N$.
We follow Refs.~\cite{Pazo2005,Ottino2016} and take natural frequencies evenly
spaced in the interval $[-1,1]$ according either to the mid-point
\begin{equation}\label{eq:mid point rule}
 \omega_i=-1+\frac{2i-1}{N} \quad i=1,\ldots,N \, ,
\end{equation}
or the end-point rule
\begin{equation}\label{eq:end point rule}
 \omega_i=-1+2\frac{i-1}{N-1} \quad i=1,\ldots,N \, ,
\end{equation}
because they allow to obtain leading-order estimates for $K_L^N$ and $r_L^N$, Eqs.~(\ref{eq:finite size locking threshold})
and (\ref{eq:finite size order parameter}) below. 
Few results we obtained with randomly but homogeneously distributed $\omega_i\in[-1,1]$ corroborate the results to be presented.
Figs. \ref{fig:K-KL scaling mid} and \ref{fig:K-KL scaling end} show numerical results for 
$N=100$ and $N=5000$ oscillators.
The data confirm the scaling predictions of 
Eqs.~(\ref{eq:critical scaling lambda2}) and (\ref{eq:scaling rc}) sufficiently close to $K_L^N$.
Thus, we report a discrepancy between the scaling of $r-r_L \sim (K-K_L^\infty)^{2/3}$
in the thermodynamic limit and our finite size 
scaling which goes like $r-r_L \sim (K-K_L^N)^{1/2}$.
Some distance away from $K_L$, one seems to recover the $N\rightarrow\infty$ behavior $\sim(K-K_L^\infty)^{2/3}$,
as is evident for $N=5000$.

\begin{figure}[htbp]
\centering
 \includegraphics[width=\columnwidth]{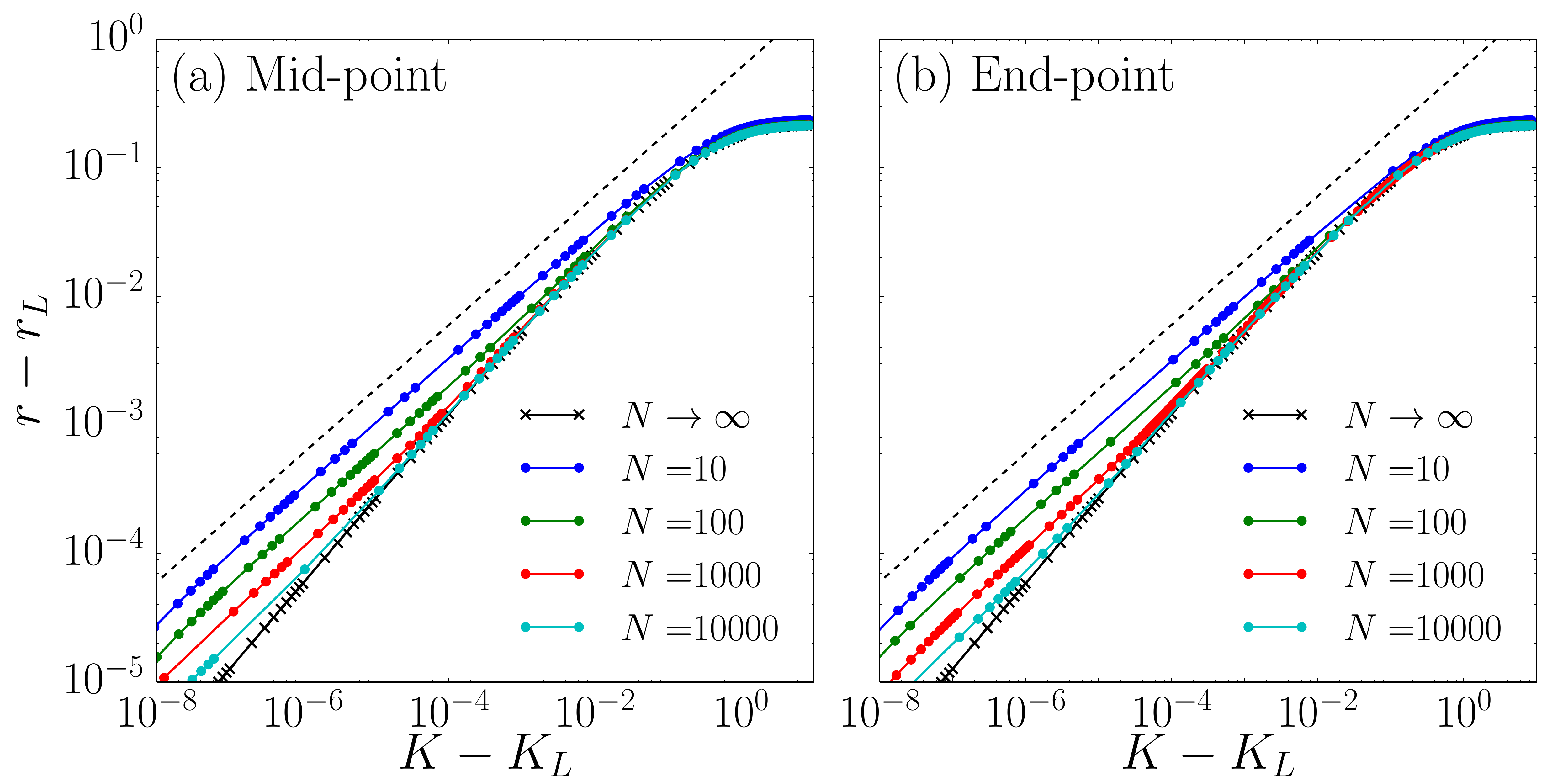}
 \caption{(Color online) Behavior of $r-r_L$ as a function of $K-K_L^N$ for different $N$.
 The coupling range above the locking threshold for which $r-r_L\sim (K-K_L^N)^{1/2}$ decreases with 
 the oscillator number. The dashed line indicating a $\sqrt{K-K_L}$ behavior is a guide to the eye.}
 \label{fig:K-KL scaling with N}
\end{figure}

\begin{figure}[htbp]
\centering
 \includegraphics[width=0.9\columnwidth]{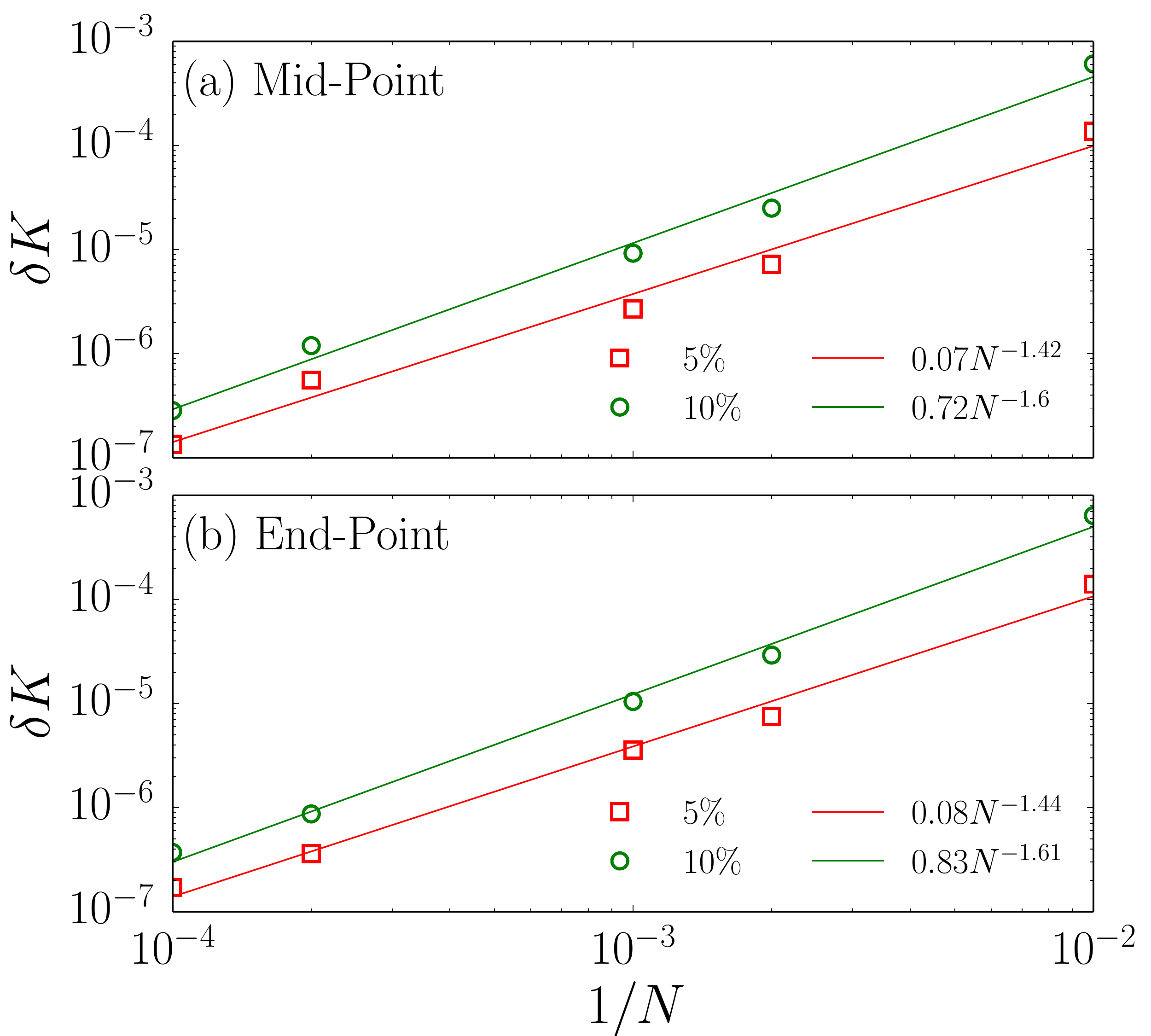}
 \caption{(Color online) Range $\delta K$ over which $r-r_L$
 deviates from our theoretical prediction Eq.~(\ref{eq:scaling rc}) by less than $5\%$ or $10\%$ 
 as a function of the number of oscillators. Solid lines are best power-law fits.
 }
 \label{fig:Region R 1/2}
\end{figure}

The above reasoning predicts that the square root behavior is valid for $K$ sufficiently close to $K_L$, but how close?
This is investigated in Figs.~\ref{fig:K-KL scaling with N} and \ref{fig:Region R 1/2} which show that the 
coupling range inside 
which the finite size scaling holds decreases with $N$. The apparent discrepancy between 
Paz\'o's \cite{Pazo2005} and our results is therefore the trademark of a crossover from finite $N$ to $N\rightarrow \infty$. 
Fig.~\ref{fig:Region R 1/2} gives the coupling range over which the numerical data obtained for $r-r_L$
deviates from our theoretical prediction, Eq.~(\ref{eq:scaling rc}), by more than $5\%$ or $10\%$,
as a function of the inverse of the oscillator number.
The observed behavior suggests that the coupling range $\delta K$ inside which $r-r_L\sim(K-K_L)^{1/2}$ 
decreases with $N$ as $\delta K\sim N^{-\alpha}$ with $\alpha\approx1.5$ 
for both mid-point and end-point frequency distributions.

While we are not able to derive analytically the value of the exponent $\alpha\approx1.5$,
we can pinpoint the origin of the crossover from $r-r_L\sim(K-K_L^N)^{1/2}$ to
$r-r_L\sim(K-K_L^\infty)^{2/3}$ as $N\rightarrow\infty$.
In our treatment above we neglected terms with $l\geq3$ in the sum over $l$ in Eq.~(\ref{eq:dLambda/dk 2}).
This is an increasingly bad approximation as the number of oscillators tends to infinity,
because then $\lambda_3(K_L^N)\rightarrow0$ as $N\rightarrow\infty$.
To show this, we recall the finite size asymptotics 
recently derived in Refs.~[\onlinecite{Pazo2005, Ottino2016}] for the mid-point and end-point frequency 
distributions, Eqs.~(\ref{eq:mid point rule}) and (\ref{eq:end point rule}).
In Ref.~[\onlinecite{Ottino2016}] Ottino-L\"{o}ffler and Strogatz obtained the finite-$N$ corrections (including
numerical prefactors) of the locking thresholds 
\begin{equation}\label{eq:finite size locking threshold}
 K_L^N=\left\{
 \begin{array}{l}
  \displaystyle \frac{4}{\pi}-\frac{64\xi}{\pi^2} N^{-3/2}+\mathcal{O}(N^{-2}) \quad \textrm{mid-point}\,,\\[3mm]
  \displaystyle \frac{4}{\pi}+\frac{4}{\pi}N^{-1}-\frac{64\xi}{\pi^2} N^{-3/2}+\mathcal{O}(N^{-2}) \quad \textrm{end-point}\,,
 \end{array}
 \right.
\end{equation}
where $\xi\approx0.093366$ is the Hurwitz zeta function evaluated at $\zeta(-1/2,C_1/2)$,
and $C_1\approx 0.605444$ is defined by $\zeta(1/2,C_1/2)=0$ with $0 \leq C_1 \leq 1$ [\onlinecite{Quinn2007, Bailey2009}].
One obtains the leading finite size corrections to the order parameter as
\begin{equation}\label{eq:finite size order parameter}
 r_L^N\approx\frac{\pi}{4}+\frac{\pi}{4}(C_1-1)N^{-1}+\mathcal{O}(N^{-3/2})\,.
\end{equation}
Despite the different scalings for the locking threshold, the asymptotic scaling of the order parameter, Eq.~(\ref{eq:finite size order parameter}), 
is the same for both frequency distributions. 
\begin{figure}[h]
\centering
 \includegraphics[width=0.9\columnwidth]{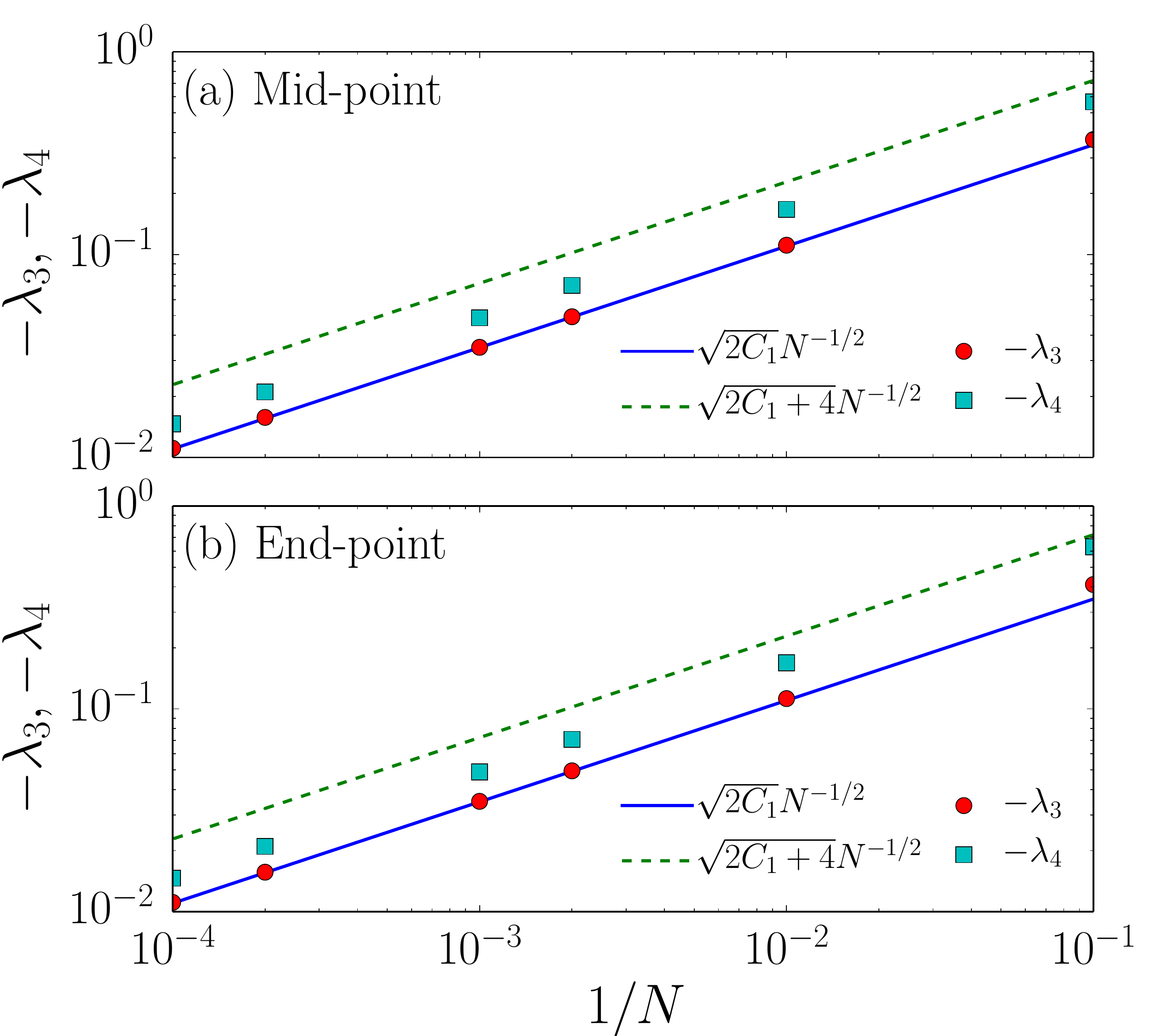}
  \caption{(Color online) 
 Second and third largest non vanishing Lyapunov exponents at the locking threshold,
 for mid-point and end-point frequency distributions (panels a) and b) respectively).
 The solid and dashed lines give the interval defined by Eq.~(\ref{eq:inequalities}).}
 \label{fig:Finite size corrections}
\end{figure}

Mirollo and Strogatz [\onlinecite{Mirollo2005}] further showed that the spectrum of the locked state for the finite size Kuramoto model 
is composed of a discrete part consisting of the eigenvalues $\lambda_1=0$ and $-\sqrt{(Kr)^2-\omega^2}\leq\lambda_2\leq0$ and
of a continuous part containing the remaining $N-2$ eigenvalues $-Kr\leq\lambda_N\leq\ldots\leq\lambda_3\leq-\sqrt{(Kr)^2-\omega^2}$
where $\omega\equiv\max_i|\omega_i|$. An additional result of Ref.~[\onlinecite{Mirollo2005}] is that for symmetric frequency distributions 
(as is the case for the mid-point and end-point rules) the Lyapunov exponents $\lambda_3$ and $\lambda_4$ can be  located even more sharply as
\begin{equation}\label{eq:location l3 and l4}
 -\sqrt{(Kr)^2-\omega_{2nd}^2} \leq\lambda_4\leq\lambda_3\leq-\sqrt{(Kr)^2-\omega^2}\,,
\end{equation}
with $\omega_{2nd}$ the second largest frequency.

At the locking threshold, $K_L^N$ and $r_L^N$, the fully locked state is marginally stable $\lambda_2=0$ and expanding the bounds of
Eq.~(\ref{eq:location l3 and l4})
in powers of $N$ using Eqs.~(\ref{eq:finite size locking threshold}) and (\ref{eq:finite size order parameter}) gives the scaling of the gap which separates
the continuous part of the spectrum from zero.
Using $\omega=1-1/N$, $\omega_{2nd}=1-3/N$ and $\omega=1$, $\omega_{2nd}=1-2/N-1$ 
respectively for mid-point and end-point rules we obtain
\begin{equation}\label{eq:inequalities}
 -\frac{\sqrt{2C_1+4}}{\sqrt{N}}\leq\lambda_4\leq\lambda_3\leq-\frac{\sqrt{2C_1}}{\sqrt{N}}\,,
\end{equation}
for both choices.
Fig.~\ref{fig:Finite size corrections} confirms numerically the scalings of $\lambda_3$ and $\lambda_4$ 
at $K_L$.

Eq.~(\ref{eq:inequalities}) then shows that at the locking threshold, $\lambda_{3,4} \sim N^{-1/2}$. 
This implies that the larger the number of oscillators, the closer to zero the continuous part 
of the spectrum will be.
Thus neglecting terms with $l\geq3$ in the sums in Eqs.~(\ref{eq:derivative order parameter3}) and (\ref{eq:dLambda/dk 2}) 
is an increasingly unjustified approximation as $N$ increases.
The exponent $1/2$ in the behaviors of $\lambda_2$ and $r-r_L$ relies on this truncation, which is
justified only in an interval $|K-K_L|<\delta K$ which is shrinking with $N$.
To recover the $2/3$ exponent obtained by Paz\'o in the continuous limit would require a resummation of all terms 
in Eqs.~(\ref{eq:derivative order parameter3}) and (\ref{eq:dLambda/dk 2}) as $N$ tends to infinity,
which we have not been able to do.

\section{Conclusion}
We investigated the scaling properties of the Kuramoto model with uniformly distributed natural frequencies
close to the synchronization threshold at finite but growing number $N$
of oscillators.
We found a non trivial behavior in that both the largest non zero Lyapunov exponent $\lambda_2$, and the order parameter $r$
of the fully locked state scale like $\lambda_2\sim(K-K_L^N)^{1/2}$ and $r-r_L^N\sim(K-K_L^N)^{1/2}$,
above the locking threshold $K_L^N$.
Our results differ from the prediction $r-r_L^\infty \sim (K-K_L^\infty)^{2/3}$ 
of Paz\'o \cite{Pazo2005} for infinitely many oscillators.
We showed that this apparent disagreement is the trademark of a crossover form finite $N$ to $N\rightarrow\infty$.
The  range of validity $\delta K$ of our result
$\lambda_2, r-r_L \sim(K-K_L^N)^{1/2}$ shrinks with $N$. We found numerically $\delta K \sim N^{-\alpha}$, with $\alpha \approx 1.5$.
Although the numerics presented in this work are for evenly spaced frequencies, 
our results remain valid for other choices of $\omega_i$'s compatible with a uniform distribution.
Our scaling predictions for $\lambda_2$ and $r-r_L$ do not depend on this choice and we checked numerically on few examples
that they remain valid for frequencies drawn randomly from a uniform distribution.

The fully locked states of the Kuramoto model have been thoroughly investigated in the limit of infinitely 
many oscillators. For the special case of uniform frequency distributions, 
long-established analytical results are known for: i) the value $K_L^\infty$ of the locking threshold, ii)
the value $r_L^\infty$ of the order parameter at phase locking and iii) the scaling behavior of the order parameter $r$ above $K_L$.
For finite $N$, however, much less is known.
Finite size corrections to the locking threshold 
and to the order parameter for frequencies uniformly distributed over the interval $[-1,1]$
have been calculated only recently~\cite{Ottino2016}. The motivation behind the present work is to investigate further the finite $N$ behavior of the 
Kuramoto close to the locking threshold.
Our manuscript complements Ref.~\cite{Ottino2016} by investigating the scalings of the largest Lyapunov exponent and of 
the order parameter above $K_L$. The observed crossover from finite to infinite $N$ and shrinking range of validity of our results 
significantly clarifies the mechanism behind the convergence to the  limit $N\rightarrow\infty$ of the Kuramoto model.

\section*{Acknowledgements}
We thank S. Strogatz for useful comments.
This work was supported by the Swiss National Science Foundation.

\bibliographystyle{apsrev}
\bibliography{bibliography}
\end{document}